\pdfoutput=1 % For ARXIV

\documentclass{article} % For LaTeX2e
\usepackage{iclr2024_conference,times}

\usepackage{hyperref}
\usepackage{url}
\usepackage[utf8]{inputenc} % allow utf-8 input
\usepackage[T1]{fontenc}    % use 8-bit T1 fonts
\usepackage{booktabs}       % professional-quality tables
\usepackage{amsfonts}       % blackboard math symbols
\usepackage{nicefrac}       % compact symbols for 1/2, etc.
\usepackage{microtype}      % microtypography
\usepackage{multirow}
\usepackage{amsmath}
\usepackage{graphicx}
\usepackage{colortbl}

\title{MedUHIP: Towards Human-In-the-Loop Medical Segmentation}

\iclrfinalcopy
\author{%
  Jiayuan Zhu \\
  University of Oxford \\
  \texttt{jiayuan.zhu@ieee.org} \\
  \And
  Junde Wu \thanks{Project Lead}\\
  University of Oxford \\
  \texttt{jundewu@ieee.org} \\
}

\begin{document}
\maketitle
\begin{abstract}
Although segmenting natural images has shown impressive performance, these techniques cannot be directly applied to medical image segmentation. Medical image segmentation is particularly complicated by inherent uncertainties. For instance, the ambiguous boundaries of tissues can lead to diverse but plausible annotations from different clinicians. These uncertainties cause significant discrepancies in clinical interpretations and impact subsequent medical interventions. Therefore, achieving quantitative segmentations from uncertain medical images becomes crucial in clinical practice. 
To address this, we propose a novel approach that integrates an \textbf{uncertainty-aware model} with \textbf{human-in-the-loop interaction}. The uncertainty-aware model proposes several plausible segmentations to address the uncertainties inherent in medical images, while the human-in-the-loop interaction iteratively modifies the segmentation under clinician supervision. This collaborative model ensures that segmentation is not solely dependent on automated techniques but is also refined through clinician expertise. As a result, our %uncertainty-aware human-in-the-loop 
approach represents a significant advancement in the field which enhances the safety of medical image segmentation. It not only offers a comprehensive solution to produce quantitative segmentation from inherent uncertain medical images, but also establishes a synergistic balance between algorithmic precision and clincian knowledge. We evaluated our method on various publicly available multi-clinician annotated datasets: REFUGE2, LIDC-IDRI and QUBIQ. Our method showcases superior segmentation capabilities, outperforming a wide range of deterministic and uncertainty-aware models. We also demonstrated that our model produced significantly better results with fewer interactions compared to previous interactive models. We will release the code to foster further research in this area.

\end{abstract}

\section{Introduction}
Medical image segmentation plays an indispensable role in disease diagnosis, prognosis monitoring and anatomy delineation \cite{mei_artificial_2020, wang_benchmark_2019, zhu_anatomynet_2019}. With the rapid development of artificial intelligence in recent decades, an enormous number of deep learning models have been applied in clinics to assist medical image segmentation \cite{litjens_survey_2017, liu_review_2021}. These medical image segmentation models often employ techniques initially developed for natural images and then modified for medical applications. Despite the apparent success of these adaptations \cite{chen_transunet_2021, ronneberger_u-net_2015, tajbakhsh_convolutional_2016}, they often overlook the unique challenges inherent in medical images \cite{hesamian_deep_2019}.

Unlike natural images, which are characterised by clear and distinct patterns, medical images typically exhibit ambiguous boundaries due to varying tissue contrast and overlapping anatomical structures \cite{hesamian_deep_2019}. Even different professional clinicians may provide different annotations for the same medical image, reflecting a unique level of uncertainty not seen in natural images. This uncertainty causes incomplete version of the ground truth, potentially resulting in unpredictable clinical outcomes \cite{cabitza_giant_2018, garcia_effect_2015}. Therefore, developing methodologies specifically to manage the uncertainties inherent in medical images becomes crucial.

A common approach to express the medical image uncertainty is generating multiple segmentations, \cite{baumgartner_phiseg_2019, kohl_probabilistic_2019, lakshminarayanan_simple_2017, rupprecht_learning_2017}, mimicing the behaviour of a group of clinicians. However, this method has a major drawback, restricting direct real-world applications. Clinicians need to review and interpret numerous predicted segmentations, which can be overwhelming and time-consuming. Besides, deciding which segmentation best captures their thought can also be challenging. In the worst-case scenario, none of the generated segmentations meet the clinician's expectations, thereby negating the potential benefits of this approach. 

A human-in-the-loop interactive model can mitigate the limitations of generating multiple segmentations by incorporating clinician feedback into the segmentation process \cite{marinov_deep_2024}. This human-in-the-loop approach allows clinicians to provide real-time corrections, so they no longer need to review and interpret numerous segmentations. Instead, the model progressively improves based on clinicians' input and moves towards clinicians' annotation, ensuring the final segmentation satisfies the clinician. Therefore, the human-in-the-loop approach enhances efficiency and robustness, making the segmentation process more practical for real-world applications \cite{wang_interactive_2018}.

In this paper, we introduce \textbf{MedUHIP}, a new paradigm that leverages both \textbf{U}ncertainty-aware models and \textbf{H}uman-\textbf{I}n-the-loo\textbf{P} interactions. MedUHIP addresses medical image uncertainty by generating multiple segmentations and fusing them into a single soft prediction. Clinicians can modify this soft prediction through human-in-the-loop interactions, allowing the model to learn their individual preferences and thus further understand the uncertainty inherent in medical images. Subsequently, the model generates a new set of segmentations that better align with the clinicians' interaction. Through this human-in-the-loop interaction process, the final prediction supported by the clinician becomes more suitable for direct clinical use.

In summary, our contributions are as follows:
\begin{itemize}
\item We introduced a novel approach, which allows iteratively modifying the prediction and captures the inherent uncertainty in the medical image. 
\item We proposed a model with Sampling Net module to learn clinician's preference based on their interaction. The Sampling Net module results a sampling space which adapts towards clinician's preference. Besides, multiple segmentations can be predicted by sampling from this space to reflect the uncertainty in medical images.
\item We compared our MedUHIP approach with various deterministic and uncertainty-aware models. It consistently achieves significantly superior results on multi-clinician annotated datasets: REFUGE2, LIDC-IDRI and QUBIQ. Additionally, MedUHIP outperforms other interaction methods with fewer iterations.
\end{itemize}

\section{Related Work}
\textbf{Medical Image Segmentation}
Deep learning is crucial for medical image segmentation as it enhances accuracy and efficiency for various diagnostic and treatment processes \cite{chan_deep_2020, wang_review_2021}. Representative models like U-Net \cite{ronneberger_u-net_2015}, %nnU-Net \cite{isensee_nnu-net_2018}, 
TransUNet \cite{chen_transunet_2021}, and SwinUNet \cite{cao_swin-unet_2021} have advanced performance by incorporating frontier computer vision techniques. U-Net is effective and simple, TransUNet captures long-range dependencies, and SwinUNet offers exceptional accuracy and scalability. 

\textbf{Uncertainty Estimation} It is notable that varying tissue contrast or clinical expertise will lead to different annotations given the same medical image \cite{sylolypavan_impact_2023}. This uncertainty inherent in medical images cannot be reduced with more data or more complex model \cite{kiureghian_aleatory_2009}. It can only be estimated by training the model to generate a range of potential predictions \cite{huang_review_2024, zou_review_2023}. Model ensembling, label sampling \cite{jensen_improving_2019}, and multi-head strategies \cite{guan_who_2018} are typical methods to address uncertainty by combining multiple model predictions. 
ProbUNet \cite{kohl_probabilistic_2018}, CM-Net \cite{zhang_disentangling_2020} and MRNet \cite{ji_learning_2021} model the the posterior distribution of model parameters or predictions explicitly to capture the uncertainty. %Bayesian inference
However, all above techniques either produce a set of predictions or require the prior knowledge of the expertise level, hindering the wide and direct clinical use. 

\textbf{Human-in-the-loop Interactive Models}
Interactive medical image segmentation is essential for improving accuracy by integrating clinicians to refine automated segmentation results. Most of previous methods \cite{luo_mideepseg_2021, sakinis_interactive_2019, wang_interactive_2018, wang_deepigeos_2019} encode clinician interactions (e.g. click, scribble, bounding box) into a distance map and integrate them into the CNN-based network. However, the choice of encoding strategies significantly impact the segmentation performance \cite{luo_mideepseg_2021}. Some approaches like Segment Anything Model (SAM, \cite{kirillov_segment_2023}) represents user interactions by positional encoding \cite{tancik_fourier_2020}, which does not require extra distance map. It maps input interactions into Fourier features to capture complex spatial and temporal relationships. Despite its extraordinary performance in natural images, a series follow-up work \cite{deng_sam-u_2023, ma_segment_2024, wu_medical_2023} in medical domain also show excellent results.

%-------------------------------------------------------------------------

\section{Methodology}
\subsection{Motivation} \label{section:motivation}
As aforementioned, the medical image uncertainty can be shown as the annotation disagreement between different clinicians. MRNet \cite{ji_learning_2021} quantitatively proved that individual clinician has consistent segmentation patterns, while the expertise levels among different clinicians vary. %Therefore, MRNet incorporate expertise level as prior knowledge to solve the medical image uncertainty problem. 
We reasonably conjecture that the interaction behaviour among individual clinicians is also consistent, but it differs among clinicians. If so, we can adaptively learn the individual clinician's preference through human-in-the-loop interaction and modify the segmentation accordingly. %and disregard the fixed expertise level prerequisite.

\begin{figure}[hbt!]
    \centering
    \includegraphics[width=0.5\linewidth]{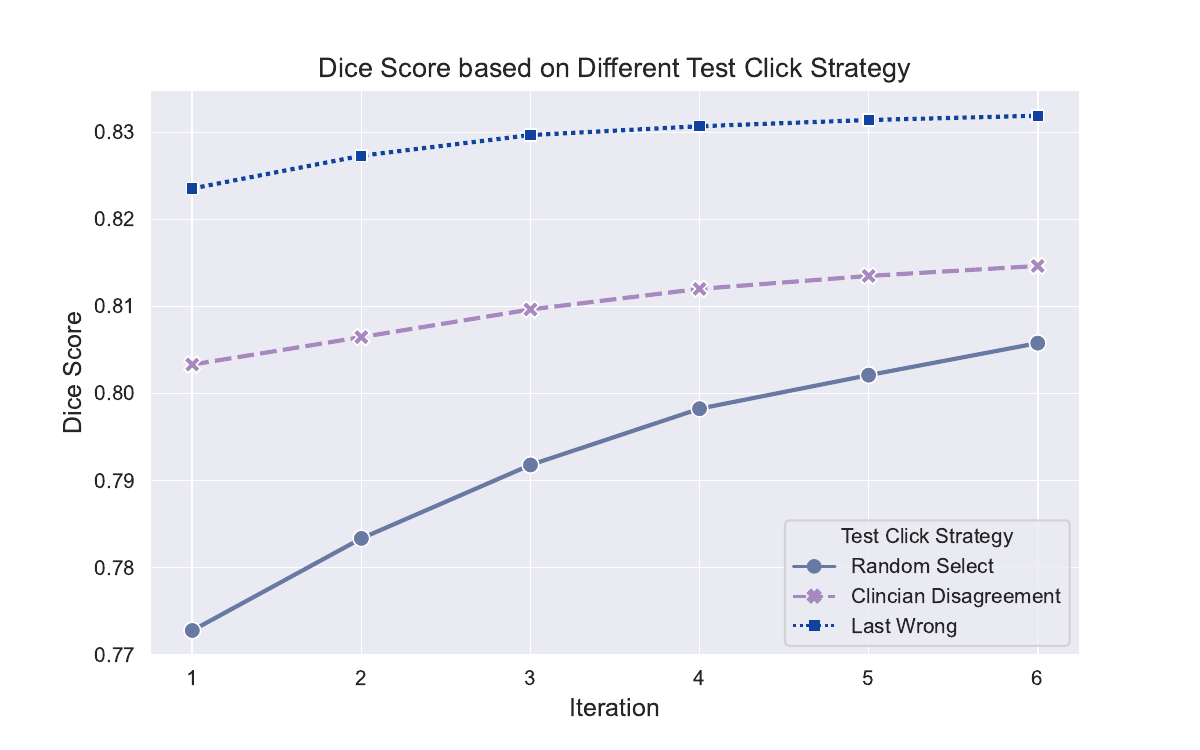}
    \caption{A preliminary experiment in testing the impact of different interaction (i.e. clicking) preferences, conducted for the optic cup segmentation on REFUGE2 test set under SAM's structure with Dice score. It indicates that the segmentation performance significantly varies across different interaction strategies, regardless of the number of clicks.}
    \label{fig:Motivation}
\end{figure}

We then conduct a preliminary experiment with optic cup segmentation on REFUGE2 dataset under SAM's \cite{kirillov_segment_2023} setting to demonstrate that clinician-specific interaction indeed impacts the segmentation performance. As we do not have the records about how each clinician provides the interaction, we mimic the behaviour through three strategies, reflecting clinicians with different expertise levels. If the clinicians have limited background knowledge, they are likely to click the image randomly. Assertive clinicians might click areas distinct to others and experienced clinicians tend to self-correct the error from last iteration. Fig. \ref{fig:Motivation} shows that the segmentation performance truly depends on the interaction strategy or individual clinician's preference. Therefore, this discovery motivated us to propose the MedUHIP model.

\subsection{Overall Framework}
\begin{figure}[hbt!]
    \centering
    \includegraphics[width=0.95\linewidth]{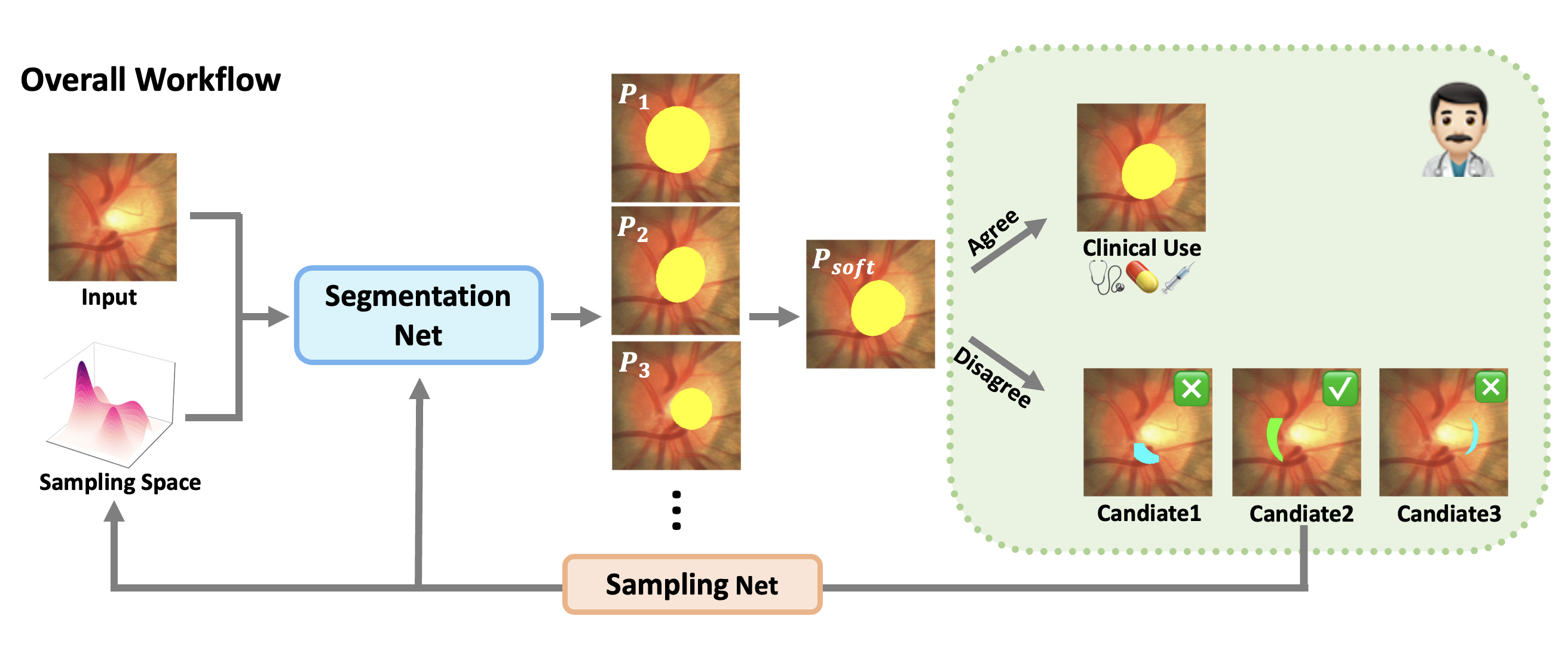}
    \caption{An overall workflow of our MedUHIP model.}
    \label{fig:Workflow}
\end{figure}

In this work, we propose the novel MedUHIP model to assist segmenting uncertain medical images through human-in-the-loop interaction. Fig. \ref{fig:Workflow} illustrates the overall framework of our MedUHIP, which contains iterative human-model interaction steps towards the final segmentation. It ensures that the final segmentation is not only predicted by the automatic model, but also approved by the clinician, which encourages direct clinical use.

In the $t^{th}$ iteration, apart from the input image $I$, we also randomly draw $N$ samples $S_1^{t}, ..., S_N^{t}$ from the sampling space $SP$ to capture the image uncertainty. 
%clinician's annotation diversity.
The combined features are sent to the Segmentation Net for automatic segmentation, generating $N$ predictions $P_1^{t}, ..., P_N^{t}$. %Segmentation Net originally adopts the frontier SAM architecture. 
These output segmentations are fused to produce a soft binary prediction $P_{soft}^{t}$, which can be direct used under clinician approval. If the clinician unsatisfies with $P_{soft}^{t}$, our model will offer K candidate regions $C_1^{t}, ..., C_K^{t}$ recommended for further improvement. These candidate regions are composed from the set of previous predicted segmentations $P_1^{t}, ..., P_N^{t}$. Afterwards, the Sampling Net modifies the sampling space $SP$ based on clinician's interaction selection, aiming to produce segmentations towards clinician's preference in the next iteration.

\begin{figure}[hbt!]
    \centering
    \includegraphics[scale=0.15]{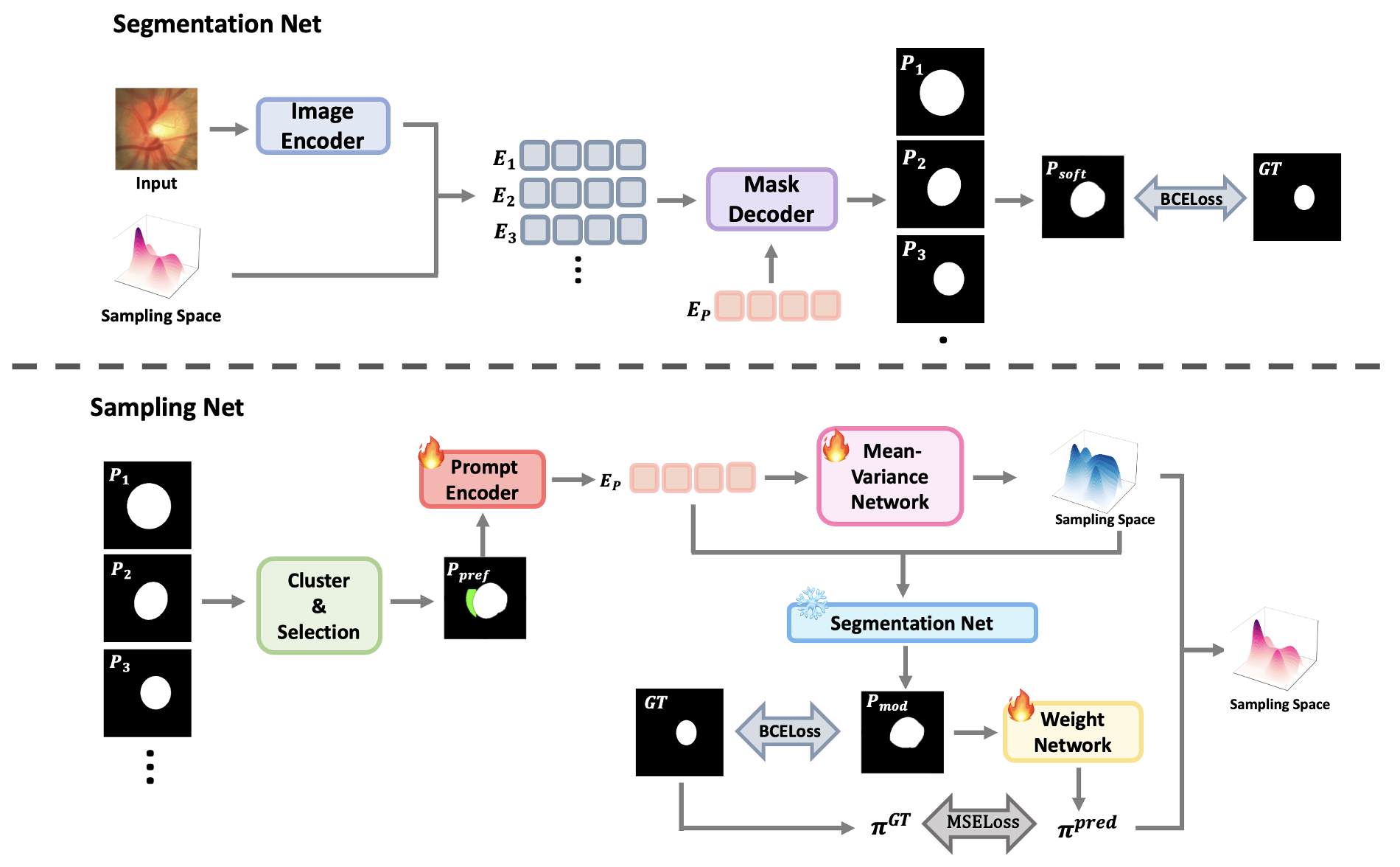} %width=0.75\linewidth
    \caption{The architecture of Segmentation Net and Sampling Net.}
    \label{fig:Network}
\end{figure}

\subsection{Segmentation Net}
Similar to SAM\cite{kirillov_segment_2023}, our segmentation net is consisted of image encoder, prompt encoder, and mask decoder. We first obtain a general image embedding $E_G^{t}$ by applying a MAE \cite{he_masked_2021} pre-trained Vision Transformer (ViT) to the input image $I$. Then we randomly draw N samples $S_1^{t}, ..., S_N^{t}$ from the sampling space $SP$. Each $S_i^{t}$ is combined with the general image embedding $E_G^{t}$ to generate a series of new image embeddings $E_1^{t}, ..., E_N^{t}$ by:
% \vspace{-2pt}
\begin{equation}
  E_i^{t} = ReLU(Conv(E_G^{t} \oplus S_i^{t})), i = 1, 2, ..., N
\end{equation}
The prompt encoder in Sampling Net maps clinician's interaction selection to 256-dimensional vectorial embedding $E_P^{t}$. It considers both the interaction location and whether it belongs to foreground or background. 

The lightweight mask decoder is a modified transformer decoder block, which integrates a dynamic mask prediction head. It utilises two-way cross-attention to enable the interaction between each image embedding $E_i^{t}$ and prompt embedding $E_P^{t}$. Afterwards, it upsamples the image embedding $E_i^{t}$, and an MLP translates the output token into a dynamic linear classifier that predicts a sequence of masks $P_1^{t}, ..., P_N^{t}$ for the input image $I$. The soft prediction $P_{soft}^{t}$ is calculated by majority vote from the sequence of masks, where $\mathbb{I}\left( \cdot \right)$ is the indicator function:
 % \vspace{-2pt}
 \begin{equation}
    P_{soft}^{t} = \mathbb{I}\left( \frac{1}{N} \sum_{i=1}^{N} P_i^{t} > \frac{1}{2} \right)
\end{equation}
We represent all parameters in Segmentation Net as $\theta_{Seg}$, and update them through the cross-entropy loss function $\mathcal{L}_{ce}$ with ground truth $GT$:
% \vspace{-2pt}
\begin{equation}
  \theta^{new}_{Seg} = \theta^{old}_{Seg} + \alpha \nabla_{\theta_{Seg}} \mathcal{L}_{ce} (P_{soft}^{t}, GT)
% \vspace{-3pt}
\end{equation}

\subsection{Sampling Net}
Given predicted masks $P_1^{t}, ..., P_N^{t}$, we first cluster them into K groups by K-means \cite{noauthor_advances_nodate} method. Then the clinician selects the potential region $P_{pref}$ which requires further modification, based on their personal preference. This information is incorporated into prompt embedding $E_p^t$ through prompt encoder.

We assume that each clinician is independent and identically distributed and their prompt embedding $E_p^t$ is conditioned on the sampling space $SP$. So we can further assume that the sampling space $SP$ is composed of M independent and identically distributed Gaussian distributions $Z_1, ..., Z_M$, each with mean $\mu_{m}$, variance $\sigma_{m}^2$ and weight $\pi_m$, where $\pi_m > 0$ and $\sum_{m=1}^{M}\pi_{m} = 1$:

\begin{equation}
  P_{pref}^{t} \mid Z_i = m \sim N(\mu_{m}, \sigma_{m}^2), m = 1, ..., M
%\vspace{-3pt}
\end{equation}

Thus, if we represent $f(P_{pref}^{t}; \mu_{m}, \sigma_{m}^2)$ as the probability density function of a Gaussian distribution with mean $\mu_{m}$ and variance $\sigma_{m}^2$, then the posterior distribution of $Z_i$ given $P_{pref}^{t}$ can be written as:
\begin{equation}
  P(Z_i = m \mid P_{pref}^{t}) = \frac{P(P_{pref}^{t} \mid Z_i = m) \cdot P(Z_i = m)}{P_{pref}^{t}} = \frac{f(P_{pref}^{t}; \mu_{m}, \sigma_{m}^2) \cdot \pi_{m}}{\sum_{m'}^{M}f(P_{pref}^{t}; \mu_{m'}, \sigma_{m'}^2) \cdot \pi_{m'}}
% \vspace{-3pt}
\end{equation}

Given the image embeddings, $\mu_{m}$ and $\sigma_{m}^2$ are trained through Mean-Variance Network, which consists of several convoluation layers and ReLU activations. In order to update all parameters in the Mean-Variance Network and prompt encoder, we compare the modified prediction $P_{mod}^{t}$ with ground truth $GT$ through the cross-entropy loss function. These parameters are represented as $\theta^{new}_{MVP}$, and importantly, all parameters in the segmentation net are frozen:
% \vspace{-2pt}
\begin{equation}
  \theta^{new}_{MVP} = \theta^{old}_{MVP} + \alpha \nabla_{\theta_{MVP}} \mathcal{L}_{ce} (P_{mod}^{t}, GT)
% \vspace{-3pt}
\end{equation}

Apart from $\theta^{old}_{MVP}$, we utilise Weight Network to update $\pi_{m}$ through the mean squared error loss function $\mathcal{L}_{mse}$ with ground truth's posterior distribution:
% \vspace{-2pt}
\begin{equation}
  \theta^{new}_{W} = \theta^{old}_{W} + \alpha \nabla_{\theta_{W}} \mathcal{L}_{mse} (P(Z_i = m \mid P_{mod}^{t}), P(Z_i = m \mid GT))
% \vspace{-3pt}
\end{equation}
where $\mathcal{L}_{mse}$ is the mean-square loss function and $\theta^{new}_{W}$ represents all parameters in the weight network. The weight network is composed by simple linear layers and ReLU activations.

\section{Experiment}
We conducted extensive experiments to verify the effectiveness of our proposed MedUHIP approach across seven multi-clinician annotated medical segmentation tasks, utilising data from various imaging modalities, including colour fundus images, CT, and MRI scans.
\subsection{Dataset}
\textbf{REFUGE2} benchmark \cite{fang_refuge2_2022} is a publicly available fundus image dataset collected for glaucoma analysis, including the optic-cup segmentation task. The fundus images are annotated by seven independent ophthalmologists, each with an average of 8 years of experience. These annotations are then reviewed by a senior specialist with over 10 years of experience in the field. REFUGE2 dataset contains 400 images for training and 400 images for testing.

\textbf{LIDC-IDRI} benchmark \cite{armato_lung_2011, clark_cancer_2013} originally consists of 3D lung CT scans with semantic segmentations of possible lung abnormalities. It comprises 1,018 lung CT scans from 1,010 patients, with manual lesion segmentations provided by four radiologists. We use a pre-processed dataset offered by \cite{kohl_hierarchical_2019}, resulting 15,096 2D CT images. After a 80-20 train-test split, our training and testing dataset contains 12,077 and 3,019 images, respectively.

\textbf{QUBIQ} benchmark \cite{li_qubiq_2024} is collected for investigating inter-clinician variability in medical image segmentation tasks. It contains one MRI brain tumour task (three annotations, 28 cases for training, 4 cases for testing); two MRI prostate-related tasks (six annotations, 48 cases for training, 7 cases for testing); one MRI brain-growth task (seven annotations, 34 cases for training, 5 cases for testing); and one CT kidney task (three annotations, 20 cases for training, 4 cases for testing).

\subsection{Experimental Setup}
Our network was implemented with the PyTorch platform and trained/tested on RTX A4000 with 32GB of memory. During training, we employed the Adam optimizer with an initial learning rate of $1e^{-4}$ and adjusted the learning rate with StepLR strategy. To ensure fair comparison, we used majority vote to train deterministic methods with multiple annotations. For the SAM-series interaction models, we randomly generated click prompt or bounding box prompt, depending on the original model settings. During the testing stage, a random set of annotations were selected, fused, and binarised. This fused binary segmentation was then used as the ground truth for evaluation. In addition, we consistently utilised vit/b as the backbone when vision transformer was involved in the models.

%-------------------------------------------------------------------------
\subsection{Experimental Result}
\subsubsection{Performance Analysis on Different Interaction Strategy}
%-------------------------------------------------------------------------
\begin{table*}[hbt]
\centering
\vspace{-10pt}
\caption{Performance Analysis by Dice Score comparison (\%) between different interaction strategies. Columns represent ground truth from combinations of 1 to 7 clinicians' annotations, comparing different interaction strategies. }
\vspace{5pt}
\resizebox{0.8\linewidth}{!}{%
\begin{tabular}{c|cccccccc|}
\hline
Interaction Strategy & 1 & 2 & 3 & 4 & 5 & 6 & 7 & Ave\\ \hline
Random Select        & 67.59  & 71.46  & 77.29  & 78.80  & 82.59  & 80.41  & 84.57 & 79.17 \\
Clinician Disagreement    & 66.48  & 71.26  & 77.76  & 81.18  & 84.44  & 83.21  & 85.34 & 80.96 \\
Last Wrong           &  66.96 & 73.30  & 78.84  & 83.31  & 86.05  & 85.72  & \textbf{88.70} &  82.96 \\ \hline
\rowcolor{cyan!40!white!20}
\textbf{MedUHIP} & \textbf{74.50}  & \textbf{79.00}  & \textbf{83.52}  & \textbf{85.13}  & \textbf{87.08}  & \textbf{88.10}  & 87.70 & \textbf{85.29}    \\ \hline
\end{tabular}%
}\label{tab:exp_interaction_strategy}
\end{table*}
%-------------------------------------------------------------------------
As demonstrated in Section \ref{section:motivation}, segmentation performance varies significantly with different interaction strategies and individual clinician's preference. We conducted quantitative experiments on the REFUGE2 test set to verify that our MedUHIP approach can generate superior segmentations regardless of the clinician's interaction strategy. Table \ref{tab:exp_interaction_strategy} presents the segmentation performance measured by Dice Score, evaluated after three human interactions. Each column represents the number of clinicians whose annotations are fused to form the ground truth, ranging from one to seven. The `Random Select', `Clinician Disagreement' and `Last Wrong' strategies follow the definition in Section \ref{section:motivation}.

Table \ref{tab:exp_interaction_strategy} shows that the `Random Select', `Clinician Disagreement' and `Last Wrong' strategies exhibit performance improvements with more fused annotations, averaging 79.17\%, 80.96\%, and 82.96\% respectively. However, our proposed method, MedUHIP, consistently outperforms these approaches. MedUHIP achieves the highest Dice Score in nearly all configurations, with an average Dice Score of 85.29\%, highlighting its robustness and effectiveness. Notably, MedUHIP performs comparably or even better than the `Last Wrong' strategy, which emulates the experienced clinicians' behaviours. Additionally, MedUHIP reaches its peak performance with a Dice Score of 88.10\% when fusing annotations from six clinicians.

%-------------------------------------------------------------------------
\subsubsection{Performance Analysis with State-of-the-arts (SOTA) Methods}
To demonstrate the advantage of the proposed MedUHIP, we compared it with the SOTA methods, classified into deterministic methods (UNet \cite{ronneberger_u-net_2015}, TransUNet \cite{chen_transunet_2021}, SwinUNet \cite{cao_swin-unet_2021}), uncertainty-based methods (Ensemble UNet, ProbUnet \cite{kohl_probabilistic_2018}, LS-Unet \cite{jensen_improving_2019}, MH-Unet \cite{guan_who_2018}, CM-Net \cite{zhang_disentangling_2020}, MRNet \cite{ji_learning_2021}), and interactive methods (SAM \cite{kirillov_segment_2023}, MedSAM \cite{ma_segment_2024}, MSA \cite{wu_medical_2023}). We also compare with the uncertainty-interactive method SAM-U \cite{deng_sam-u_2023} with both SAM and MedSAM backbone. For the interactive methods, we include the results after one interaction and three interactions.

Table \ref{tab:SOTA result} provides a quantitative performance analysis by comparing Dice Scores across multiple datasets. As shown in Table \ref{tab:SOTA result}, our proposed MedUHIP after three interactions consistently achieves superior performance compared to other approaches, achieving an average Dice Score of 89.68\%. The performance improvement is especially prominent in the LIDC segmentation task, with an increase of $\sim$ 20\% over the current SOTA methods. These results underscore MedUHIP's effectiveness and robustness across diverse medical image tasks. In addition, even the performance after one interaction is considerable better than SOTA methods, highlighting its potential to generate superior segmentation with minimal user interaction.

\begin{figure}[hbt!]
    \centering
    \includegraphics[scale=0.5]{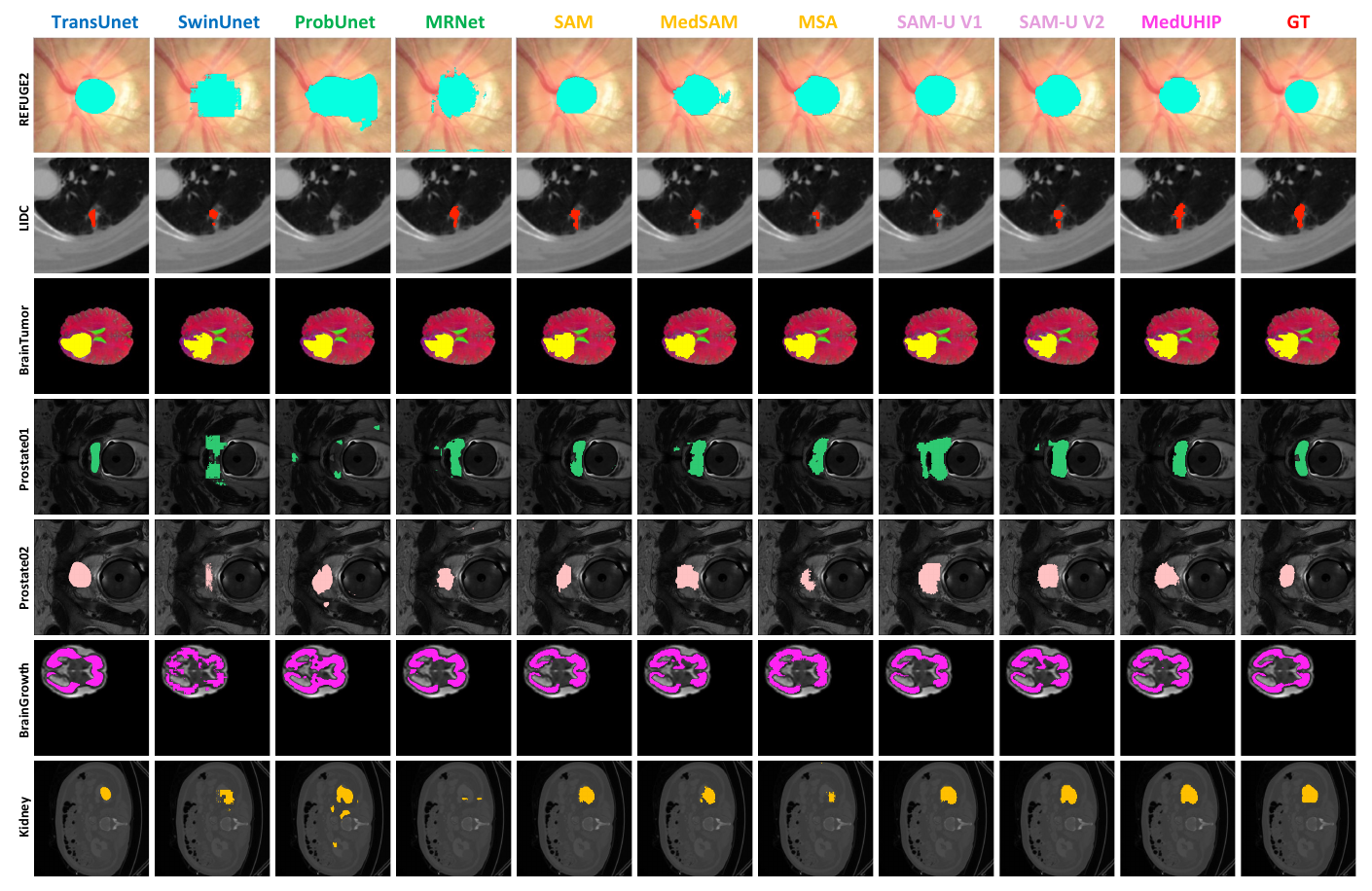
    } %width=0.75\linewidth
    \caption{Visualisation results produced by determinstic models, uncertainty-based models, interactive models, our MedUHIP method and the ground truth.}
    \label{fig:Visualisation}
\end{figure}

Fig. \ref{fig:Visualisation} illustrates the visualisation results produced by our MedUHIP in comparison with other SOTA methods. We present the segmentation after six interactions for interaction models. The ground truth is combined with randomly selected clinicians. It is evident that our model has better capability to adapt to the variability introduced by different clinicians, especially at the boundary regions.

%-------------------------------------------------------------------------
\begin{table*}[hbt!]
\centering
% \vspace{5pt}
\caption{Performance Analysis by Dice Score comparison (\%) between deterministic models, uncertainty-based models and interactive models. }
\vspace{5pt}
\resizebox{0.95\linewidth}{!}{
\begin{tabular}{c|cccccccc}
\hline
Methods      & REFUGE2  & LIDC     & BrainTumor   & Prostate1 & Prostate2 &BrainGrowth                & Kidney   & Ave            \\ \hline
UNet  \cite{ronneberger_u-net_2015}     & 68.94       & 62.99        & 87.30                     & 83.89               & 77.22        & 62.02                          & 82.40       & 74.96                                     \\
TransUNet  \cite{chen_transunet_2021}  & 80.83       & 64.09        & 90.14                     & 83.35               & 68.34        & 86.58                          & 52.99       & 75.19                                    \\ 
SwinUNet \cite{cao_swin-unet_2021}  & 78.67       & 59.45        & 91.23                     & 82.02               & 74.19        & 74.88                          & 69.41       & 75.69                                     \\ 
\hline
\begin{tabular}[c]{@{}c@{}} Ensemble UNet\end{tabular}   
             & 70.75       & 63.84        & 90.56                     & 85.27               & 79.07        & 71.69             & 89.30       & 78.64                                     \\  
ProbUnet \cite{kohl_probabilistic_2018}     & 68.93       & 48.52        & 89.02                     & 72.13               & 66.84        & 75.59                         & 75.73       & 70.96                                     \\
LS-Unet \cite{jensen_improving_2019}      & 73.32       & 62.05        & 90.89                    & 87.92               & 81.59        & 85.63                          & 72.31       & 79.10                                     \\ 
MH-Unet \cite{guan_who_2018}      & 72.33       & 62.60        & 86.74                     & 87.03               & 75.61        & 83.54                          & 73.44       & 77.32                                     \\  
MRNet \cite{ji_learning_2021}       & 80.56       & 63.29        & 85.84                    & 87.55               & 70.82        & 84.41                          & 61.30       & 76.25                                     \\ 
\hline
\begin{tabular}[c]{@{}c@{}} SAM\textsuperscript{3} \cite{kirillov_segment_2023}\end{tabular}  
       & 82.59       & 66.68        & 91.55                    & 92.82               & 77.04        & 86.63                          & 85.72       & 83.29                                     \\
MedSAM\textsuperscript{3}   \cite{ma_segment_2024}  & 82.34       & 68.42        & 92.67                     & 89.69               & 74.70        & 85.91                          & 78.02      & 81.68                                     \\ 
MSA\textsuperscript{3} \cite{wu_medical_2023} & 83.03       & 66.88        & 88.16                    & 89.06               & 68.94        & 80.62                         & 25.29       & 71.71                                     \\ \hline
SAM-U\textsuperscript{3} (SAM backbone) \cite{deng_sam-u_2023} & 82.45       & 62.24        & 92.67                     & 81.46               & 66.56        & 87.79                          & 89.50       & 80.38                                     \\
SAM-U\textsuperscript{3} (MedSAM backbone) \cite{deng_sam-u_2023} & 80.66       & 64.82        & 93.11                     & 91.89               & 72.91        & 87.51                          & 90.74       & 83.09                                     \\\hline
\rowcolor{cyan!40!white!20}
\textbf{MedUHIP\textsuperscript{1}}             & \textbf{83.47}       & \textbf{88.07}        & \textbf{94.29}                    & \emph{\textbf{93.12}}               & \textbf{83.34}        & \textbf{88.14}                          & \textbf{94.08}       & \textbf{89.22}                                     \\ \hline
\begin{tabular}[c]{@{}c@{}} SAM\textsuperscript{3}\end{tabular}  
       & 82.61       & 66.71        & 92.14                     & 92.72               & 77.54        & 86.58                          & 90.43       & 84.10                                     \\ 
MedSAM\textsuperscript{3}     & 82.13       & 68.45        & 93.26                   & 90.05               &  73.81        & 86.09                          & 79.88       & 81.95                                  \\ 
MSA\textsuperscript{3}& 83.08       & 66.87        & 91.25                  & 90.22               & 71.34        & 81.87                          & 46.76       & 75.91                                     \\ \hline
SAM-U\textsuperscript{3} (SAM backbone) & 82.10       & 62.84        & 92.31                     & 81.79               & 66.74        & 87.84                          & 89.24       & 80.40                                     \\ 
SAM-U\textsuperscript{3} (MedSAM backbone) & 80.54       & 65.44        & 92.40                     & 90.00               & 73.17        & 87.87                          & 91.35       & 82.96                                     \\ \hline
\rowcolor{cyan!40!white!20}
\textbf{MedUHIP\textsuperscript{3}}             & \emph{\textbf{85.42}}       & \emph{\textbf{88.56}}        & \emph{\textbf{94.31}}                   & \textbf{92.97}               & \emph{\textbf{84.05}}        & \emph{\textbf{88.18}}                          & \emph{\textbf{94.26}}       & \emph{\textbf{89.68}}                                     \\ \hline

\end{tabular}
}\label{tab:SOTA result}
\end{table*}
%-------------------------------------------------------------------------

%-------------------------------------------------------------------------

\subsubsection{Threshold Analysis on Different Interactive Models}
We conducted a threshold analysis to quantify the number of interactions required to reach specific Dice Scores across various interactive models. We ran all models with at most 6 iterations, and assigned the click number to 10 if the image failed to reach the specific Dice Score. Our proposed method, MedUHIP, demonstrated superior efficiency in requiring fewer interactions to achieve high-performance segmentation results compared to other methods. 

For the REFUGE2 dataset, MedUHIP only requires 2.72 interactions on average to achieve 80\% Dice Score, compared to SAM's 3.65 and MedSAM's 3.67 interactions. Similarly, for the LIDC dataset, MedUHIP necessitates just 2.24 interactions, whereas the closest competitor, SAM-U with MedSAM backbone requires 3.86 interactions. Although MedUHIP does not always require the fewest clicks, it remains highly competitive. For example, MedUHIP achieves 2.42 interactions in the Prostate1 task, which is fewer than most models, but slightly more than SAM at 1.57 interactions.

%-------------------------------------------------------------------------
\begin{table*}[hbt!]
\centering
% \vspace{5pt}
\caption{Threshold Analysis by Number of Interactions to achieve specific Dice Score between interactive models. If the model fails to achieve the Dice Score in 6 interactions, it is assigned with 10 interactions.}
\vspace{5pt}
\resizebox{0.8\linewidth}{!}{%
\begin{tabular}{c|ccccccc}
\hline
Methods         & REFUGE2 & LIDC & BrainTumor & Prostate1 & Prostate2 & BrainGrowth & Kidney  \\
              & NoC$_{80}$            &NoC$_{70}$             &NoC$_{92}$             &NoC$_{90}$           &NoC$_{80}$               &NoC$_{88}$               &NoC$_{90}$                        \\ \hline
SAM     &  3.65 &  3.94  &   2.00     &  \textbf{1.57}  &  \textbf{2.28} & 4.60  &  4.25 \\
MedSAM  &  3.67 &  3.69  &   \textbf{1.00} &  3.57  &  3.57 & 8.20  &  10.00 \\
MSA  &  3.54 & 3.88   &  5.00  & 3.00  & 3.57  &  10.00  & 10.00  \\
SAM-U (SAM backbone)  &  3.98 &  4.18 & 4.00  &  8.71  & 6.14 & 4.60 & 3.75      \\
SAM-U (MedSAM backbone) &  4.53 &  3.86 & 4.00 & 3.57  & 3.57 & 3.20 & 3.50     \\ \hline
\rowcolor{cyan!40!white!20}
\textbf{MedUHIP}  &\textbf{2.72} & \textbf{2.24} & \textbf{1.00}  & 2.42 & \textbf{2.28}  & \textbf{2.80}  & \textbf{3.25} \\ \hline
\end{tabular}%
}\label{tab:amos}
\end{table*}
%-------------------------------------------------------------------------

%-------------------------------------------------------------------------

\subsubsection{Sampling Space Analysis}
\begin{figure}[hbt!]
    \centering
    \includegraphics[width=0.6\linewidth]{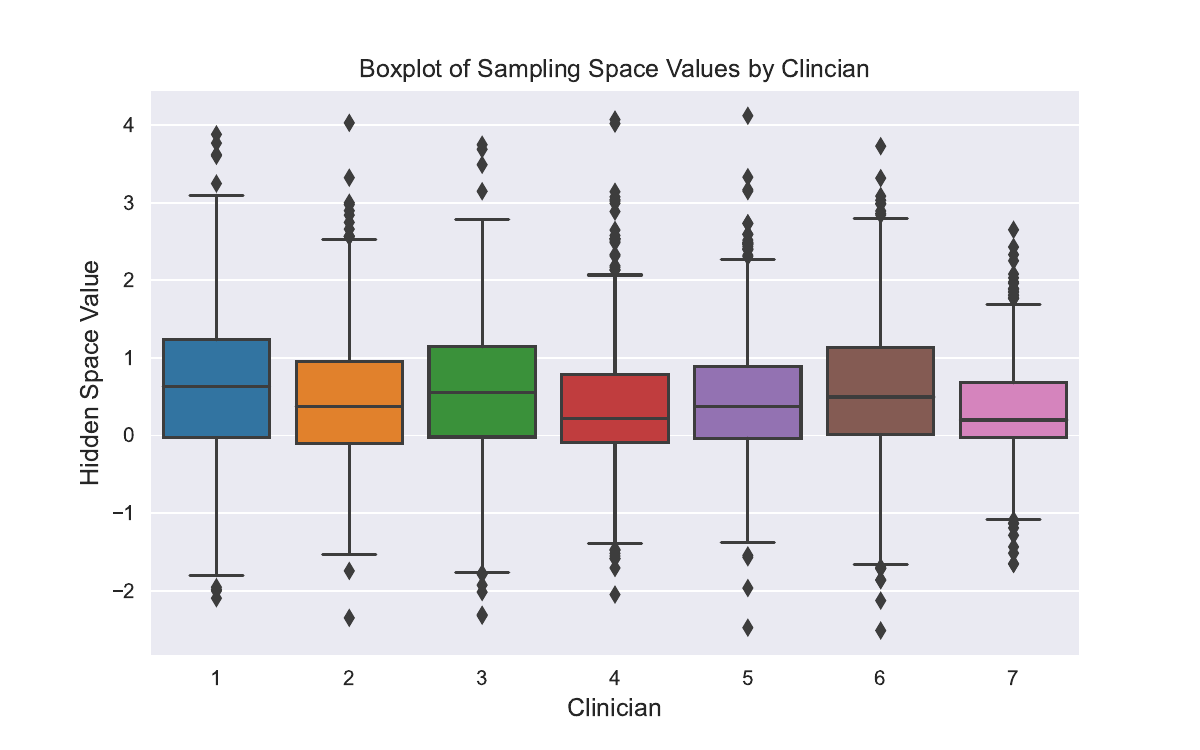}
    %[scale=0.4]
    \caption{Boxplot of sampling space values based on different clinicians. We draw samples with the mean, variance, weight after six interactions for each clinician.}
    \label{fig:Hidden_space}
\end{figure}

We carried out a sampling space analysis to demonstrate that the sampling space captures clinician interaction preference. Fig. \ref{fig:Hidden_space} represents the hidden space values for different clinicians, sampled from a Gaussian mixture model based on the mean, variance, and weight after six interactions. Each clinician on the horizontal axis has a corresponding boxplot showing the distribution of hidden space values. The similar median values across clinicians suggest that their decisions are roughly similar, while variations indicate personal preferences. It highlights that our model's hidden space can capture each clinician's unique interaction preferences.

%-------------------------------------------------------------------------
\subsubsection{Ablation Studies}
In this section, we performed an ablation study on each component of our proposed MedUHIP model, including random sampling from the hidden space, updating Gaussian distributions' mean and variance in the hidden space, and updating the weights of Gaussian distributions in the hidden space. The ablation analysis is presented in Table \ref{tab:ablation}, with segmentation performance evaluated by Dice Score for the REFUGE2 and Kidney datasets.

When only sampling from hidden space without calibrating mean, variance, and weight, the Dice Score is 80.29\% for REFUGE2 and 90.05\% for the Kidney dataset. Training the distributions' mean and variance in the hidden space boosts segmentation performance to 84.12\% and 92.06\% for REFUGE2 and Kidney, respectively. Including training distribution weight alone improves the Dice Score to 82.87\% for REFUGE2 and 92.29\% for Kidney.

Finally, combining all three components which results to calibrating distribution mean, variance, and weight before sampling from the hidden space yields the highest performance, with Dice Scores of 85.42\% for REFUGE2 and 94.26\% for the Kidney dataset. This demonstrates the complementary benefits of each component, highlighting their significance in achieving optimal segmentation performance.

%-------------------------------------------------------------------------
\begin{table*}[hbt!]
\centering
% \vspace{5pt}
\caption{Ablation analysis on REFUGE2 and Kidney dataset }
\vspace{5pt}
\resizebox{0.6\linewidth}{!}{%
\begin{tabular}{ccccc}
\hline
Sampling  & Mean-Variance & Weight & REFUGE2 & Kidney \\ \hline
\checkmark &   &   &   80.29 & 90.05  \\
\checkmark & \checkmark  &          &  84.12 & 92.06  \\
\checkmark &           & \checkmark &  82.87 & 92.29\\
\rowcolor{cyan!40!white!20}
\checkmark &    \checkmark   & \checkmark & \textbf{85.42} &  \textbf{94.26} \\ \hline
\end{tabular}%
}\label{tab:ablation}
\end{table*}
%-------------------------------------------------------------------------

\section{Conclusion}
In this work, we introduce MedUHIP, a novel paradigm that integrates an uncertainty-aware model with human-in-the-loop interaction to achieve quantitative segmentations of uncertain medical images under clinician supervision. Our approach feature a Segmentation Net, which generates multiple plausible predictions by sampling to address inherent uncertainties. We then incorporated the clinician interactions to quickly calibrate these predictions toward their preference using the Sampling Net. Extensive empirical experiments have demonstrated the superior performance of MedUHIP across a wide range of medical image segmentation tasks and diverse image modalities, all with fewer interactions. This approach shows great potential for practical implementation in clinical settings.

%-------------------------------------------------------------------------

\clearpage

\bibliography{iclr2024_conference}
\bibliographystyle{iclr2024_conference}

\end{document}